\documentclass[10pt,conference]{IEEEtran}
\IEEEoverridecommandlockouts
\usepackage{cite}
\usepackage{amsmath,amssymb,amsfonts}
\usepackage{algorithmic}
\usepackage{graphicx}
\usepackage{textcomp}
\usepackage{xcolor}
\usepackage{enumitem}
\newcommand{\wcag}{WCAG}
\newcommand{\wcagone}{WCAG 1.0}
\newcommand{\wcagtwo}{WCAG 2.0}
\newcommand{\wcagtwoone}{WCAG 2.1}
\newcommand{\wcagtwotwo}{WCAG 2.2}
\newcommand{\myparagraph}[1]{\vspace{0.1cm}\textbf{#1}}
\newcommand{\etal}{et~al.}
\newcommand{\eg}{e.g.,~}
\def\BibTeX{{\rm B\kern-.05em{\sc i\kern-.025em b}\kern-.08em
    T\kern-.1667em\lower.7ex\hbox{E}\kern-.125emX}}
    \makeatletter
\newcommand{\linebreakand}{%
  \end{@IEEEauthorhalign}
  \hfill\mbox{}\par
  \mbox{}\hfill\begin{@IEEEauthorhalign}
}
\makeatother
\begin{document}

\title{Non-Western Perspectives on Web Inclusivity: A  Study of Accessibility Practices in the Global South}
\author{\IEEEauthorblockN{Masudul Hasan Masud Bhuiyan}
\IEEEauthorblockA{\textit{CISPA Helmholtz Center for Information Security} \\
Saarbrücken, Germany \\
masudul.bhuiyan@cispa.de}
\and
\IEEEauthorblockN{Matteo Varvello}
\IEEEauthorblockA{\textit{Nokia Bell Labs} \\
New Jersey, USA \\
matteo.varvello@nokia.com}
\linebreakand
\IEEEauthorblockN{Cristian-Alexandru Staicu}
\IEEEauthorblockA{\textit{CISPA Helmholtz Center for Information Security} \\
Saarbrücken, Germany \\
staicu@cispa.de}
\and
\IEEEauthorblockN{Yasir Zaki}
\IEEEauthorblockA{\textit{New York University Abu Dhabi} \\
Abu Dhabi, UAE\\
yasir.zaki@nyu.edu}
}
\definecolor{grassgreen}{RGB}{0,128,0}
\newif\ifcomment
\commenttrue
\commentfalse

\ifcomment
    \newcounter{MVNumberOfComments}
    \newcommand{\mvnote}[1]{\textcolor{blue}{\small \bf [MV\#\arabic{MVNumberOfComments}\stepcounter{MVNumberOfComments}: #1]}} 
    \newcounter{YZNumberOfComments}
    \newcommand{\yznote}[1]{\textcolor{orange}{\small \bf [YZ\#\arabic{YZNumberOfComments}\stepcounter{YZNumberOfComments}: #1]}}
    \newcounter{MHNumberOfComments}
    \newcommand{\mhnote}[1]{\textcolor{grassgreen}{\small \bf [MH\#\arabic{MHNumberOfComments}\stepcounter{MHNumberOfComments}: #1]}} 
    \newcounter{CSNumberOfComments}
    \newcommand{\csnote}[1]{\textcolor{purple}{\small \bf [CS\#\arabic{CSNumberOfComments}\stepcounter{CSNumberOfComments}: #1]}} 
    
\else
    \newcommand\mvnote[1]{}
    \newcommand\yznote[1]{}
    \newcommand{\mhnote}[1]{}
    \newcommand{\csnote}[1]{}
\fi

\maketitle

\begin{abstract}
The Global South faces unique challenges in achieving digital inclusion due to a heavy reliance on mobile devices for internet access and the prevalence of slow or unreliable networks. While numerous studies have investigated web accessibility within specific sectors such as education, healthcare, and government services, these efforts have been largely constrained to individual countries or narrow contexts, leaving a critical gap in cross-regional, large-scale analysis. This paper addresses this gap by conducting the first large-scale comparative study of mobile web accessibility across the Global South. 
In this work, we evaluate 100,000 websites from 10 countries in the Global South to provide a comprehensive understanding of accessibility practices in these regions. Our findings reveal that websites from countries with strict accessibility regulations and enforcement tend to adhere better to Web Content Accessibility Guidelines (WCAG)  guidelines. However, accessibility violations impact different disability groups in varying ways. Blind and low-vision individuals in the Global South are disproportionately affected, as only 40\% of the evaluated websites meet critical accessibility guidelines. This significant shortfall is largely due to developers frequently neglecting to implement valid alt text for images and ARIA descriptions, which are essential specification mechanisms in the HTML standard for the effective operation of screen readers. \csnote{We are missing a clear takeaway here: what did we learn from this? What is the next step}
\end{abstract}


\section{Introduction}
The Global South refers to a broad set of countries primarily located in Africa, Asia, Latin America, and Oceania. These regions are characterized by developing economies, diverse cultures, and a large proportion of the world's population. Collectively, 
they represent over 6 billion people 
or approximately 80\% of the world’s population~\cite{DESA_2022,world_economics}. These countries are marked by their varied economic, social, and cultural landscapes, but many share common challenges in economic development, technological infrastructure, and access to essential resources. \csnote{This last sentence does not add much to the argumentation}\mvnote{agreed, i changed it} Although 
economic conditions often limit access to 
internet access, mobile technology has become a transformative force. 
For many in the Global South, mobile devices serve as the primary, and often only, means of accessing the internet, as mobile technology is generally more affordable and accessible than traditional desktop computing~\cite{UNDP_2022, GSMA_2022}.

\csnote{A bit too abrupt the jump to accessibility here. I would maybe insist on increase adoption, but possible low inclusivity due to limited financial resources to go beyond the bare minimum/functionality.} \mvnote{I have restructured a bit, merging and moving things around. I am ok with it now}


With the increasing penetration of mobile internet in the Global South, web accessibility has emerged as a critical issue. Web accessibility refers to the practice of creating inclusive digital environments, enabling people of all abilities to interact with online content. Inaccessible websites limit users’ access to essential information, digital services, and economic or social opportunities, deepening existing inequalities~\cite{noauthor_draft_nodate, noauthor_towards_nodate}. Mobile platforms face unique accessibility challenges due to smaller screen sizes, touch-based interactions, and device variability. These challenges require dedicated solutions that go beyond traditional desktop accessibility practices~\cite{initiative_wai_web_nodate, noauthor_what_nodate, muff_desktop_2017}.

Recognizing these differences, the World Wide Web Consortium (W3C) proposed updates to the Web Content Accessibility Guidelines (WCAG) in versions 2.1 and 2.2, focusing on mobile-specific issues such as small touch targets, adaptable layouts, and gestures~\cite{w3c_mobile}. However, 
the adoption of such guidelines remains questionable, as many developers are either unaware of mobile-specific requirements or do not prioritize them~\cite{initiative_wai_web_nodate, noauthor_access_nodate}. Moreover, the desktop and mobile versions of the same website often present different accessibility issues~\cite{noauthor_what_nodate, noauthor_mobile_2024}, as developers need to take additional steps to make mobile versions fully accessible. This gap in accessibility practices further underscores the need for targeted efforts to address the unique barriers mobile web users face.

\csnote{I would like to see a short paragraph about accessibility on the web: cite old papers that say the adoption is slow or the CHI paper I shared with you that says that alt descriptions is missing in a third of the images.}\mhnote{added}

Although there has been a growing body of research focused on WCAG, much of this work has concentrated on specific website categories or has been limited to single-country studies. 
Previous studies have examined accessibility in the public sector~\cite{isamova2017usability, nso2024web, mashau2023accessibility, lewthwaite2014web} or government websites within individual countries~\cite{paul2023accessibility, adepoju2016accessibility, akgul2016web, ismailova2017survey}, providing insights into specific sectors but often overlooking a cross-national perspective. This narrow scope has limited the generalizability of findings and has created a knowledge gap regarding how accessibility practices vary across the Global South~\cite{abanumy2005eGovernment, mtebe2017accessibility}. This lack of regional and cross-national analysis restricts the insights needed to support broad-based accessibility initiatives in the Global South and hampers efforts to develop inclusive policies and practices that address the unique needs of these regions~\cite{barbareschi2021challenges}.

To address this gap, 
we conduct the first large-scale comparative analysis of mobile web accessibility across the Global South, evaluating 100,000 websites from 10 countries. Our study is automated using Lighthouse, an audit tool that enables accessibility assessments aligned with the WCAG standards~\cite{W3C_WCAG2.1_2018}. 
This approach not only fills a critical gap in the literature but also lays a foundation for actionable improvements in web accessibility practices across the Global South.
\csnote{Say what you aim to do before explaining how you do it.}\mhnote{updated}
This paper makes three primary contributions to the understanding of web accessibility in the Global South: \mvnote{this is a bit weak and redundant, revisit after reading the rest of the paper}

\begin{enumerate}[left=0pt]
    
    

    \item \textbf{Comprehensive Analysis}: This study examines mobile web accessibility issues across 10 countries in the Global South, providing a broad perspective on practices and challenges in diverse contexts.

    \item \textbf{Identification of Common Challenges}: Our analysis highlights recurring issues like inadequate touch targets, low-contrast elements, and missing alt text, emphasizing systemic barriers requiring urgent attention.
    
    \item \textbf{Impact on Diverse Disability Groups}: Our findings highlight how accessibility violations disproportionately affect different disability groups in the Global South, such as blind users (missing alt text, \texttt{meta-viewport} issues) and users with motor impairments (small touch targets).

\end{enumerate}

\section{Related Work}
Accessibility research has grown recently but remains focused on the Global North, despite over 80\% of people with disabilities living in the Global South~\cite{pal2016accessibility, heeks2022digital}. 
Studies on mobile phone use in these regions reveal a heavy reliance on social networks and community support 
to address barriers like limited digital fluency and inaccessible designs~\cite{barbareschi2021challenges}. Previous research also explores digitalization~\cite{pal2016accessibility, heeks2022digital}, accessibility gaps~\cite{mashau2023accessibility, nso2024web}, and usability~\cite{mtebe2017accessibility, isamova2017usability} issues in mobile and web platforms. Sector-specific studies, such as those on university websites in Kyrgyzstan, Africa, Asia, and Latin America, reveal significant accessibility gaps hindering navigation for users with disabilities~\cite{isamova2017usability, mashau2023accessibility}.

Government websites are a major focus in accessibility research due to their role in public services. Studies from Tanzania, Turkey, Saudi Arabia, Oman, and Nigeria highlight widespread failures to meet WCAG standards, often due to limited awareness, resources, or policy enforcement~\cite{mtebe2017accessibility, akgul2016web, abanumy2005eGovernment, adepoju2016accessibility}. Similarly, healthcare websites frequently lack accessibility, creating barriers to vital medical information~\cite{fernandes2023readily, sarita2022accessibility}. Alajarmeh~\etal~\cite{alajarmeh2022evaluating} found low WCAG 2.0 compliance in healthcare websites across 25 countries. Italian tourism portals ~\cite{dattolo2016web} also suffer from design issues affecting users with autism spectrum disorders. While valuable, these studies often focus on specific countries or sectors. Our research bridges this gap by analyzing accessibility practices across multiple sectors and countries in the Global South, offering a broader perspective for targeted interventions and policies.

\section{Methodology}
Our methodology involves identifying and analyzing 100,000 popular websites from 10 Global South countries, focusing on local top-level domain (TLD) or in-country hosting. Accessibility is evaluated using Google Lighthouse. %

\subsection{Country and Website Selection}
We select 10 countries from the Global South based on population size and web data availability in the CrUX database~\cite{google_crux}. These countries—China, India, Indonesia, Pakistan, Brazil, Nigeria, Bangladesh, Mexico, the Philippines, and Vietnam—cover approximately 4.19 billion people, or 51.7\% of the global population. For each country, we analyze 10,000 websites, identified through a combination of country code top-level domains (ccTLDs) or registration data verified using \texttt{python-whois}~\cite{noauthor_python-whois_nodate} and the \texttt{who.is} API~\cite{noauthor_whois_nodate}. This method, applied to CrUX's top-ranked websites for each country, ensures a diverse sample while prioritizing locally relevant websites and minimizing the effect of global platforms, capturing the unique digital ecosystems of these nations.

\subsection{Accessibility Evaluation}

To assess each website’s accessibility, we use Google Lighthouse, an automated tool that evaluates web content against the Web Content Accessibility Guidelines (WCAG). Lighthouse calculates an Accessibility score as a weighted average of audits, informed by axe-core~\cite{noauthor_axe-coredocrule-descriptionsmd_nodate}, an open-source engine that analyzes user impact and identifies compliance issues. Our evaluation specifically targets \wcagtwotwo, which is organized around four key principles: Perceivable, Operable, Understandable, and Robust (POUR). These principles ensure accessibility for diverse user needs, including those with disabilities, across various devices. Lighthouse evaluates several critical aspects of accessible mobile design, encompassing:

\begin{enumerate}[left=0pt]
    \item \textbf{Color Contrast:} Verifies adequate contrast for readability by users with visual impairments.
    \item \textbf{Touch Targets:}  Ensures interactive elements are mobile-friendly, helping users with motor impairments.
    \item \textbf{Alternative Text:} Checks for descriptive alt text on images for screen reader users.
    \item \textbf{Form Labels:} Confirms proper labeling of form fields for easier navigation with assistive technologies.
    \item \textbf{ARIA Descriptions:} Ensures Accessible Rich Internet Applications (ARIA) descriptions are properly implemented for assistive technologies.
\end{enumerate}

These evaluations are compiled into an aggregated \textit{accessibility score} ranging from 0 to 100, providing a quantitative measure of each website’s overall accessibility performance. 

\subsection{Alignment with Accessibility Standards}
Our approach adheres to \wcagtwotwo, with a focus on Levels A and AA, which set essential and enhanced accessibility standards for usability. Developed from \wcagone, \wcagtwo~introduced adaptable principles and technology-neutral recommendations, later expanded in \wcagtwoone~to support mobile accessibility. By aligning with these standards, our analysis reflects widely accepted expectations for accessible design. By applying Lighthouse’s WCAG-aligned evaluations on popular websites in each country, this methodology offers a robust cross-country perspective on mobile web accessibility practices in the Global South. This approach highlights the strengths and areas for improvement within each country’s digital landscape, identifying common accessibility challenges and opportunities for inclusive design across the region. Our analysis provides valuable insights into how accessibility is implemented on high-traffic mobile websites in these countries, supporting efforts to make the web more inclusive and user-friendly for all. 

\section{Results}
We conduct the accessibility audit using Google Lighthouse \cite{noauthor_lighthouse_nodate}, with a 30-second timeout per webpage. To ensure mobile rendering, audits emulate an iPhone X (375x812 resolution, scale factor of 3). Our software runs on university servers in Germany to reduce bias from commercial cloud IPs~\cite{jueckstock2021towards}.

\subsection{Accessibility Scores Across Regions}
\label{sec:accessibility_scores_regions}


\begin{table}
\centering
\setlength{\tabcolsep}{3pt}
\caption{Average number of accessibility violations per website by country}
\vspace{-2mm}
\begin{tabular}{ccc|ccc}
\hline
\textbf{Country} & \textbf{All Issue} & \textbf{Critical} & \textbf{Country} & \textbf{All Issue} & \textbf{Critical} \\
\hline
Bangladesh (bd) & 5.48 & 1.29 & India (in) & 4.40 & 1.06 \\
Vietnam (vn) & 4.92 & 1.33 & Mexico (mx) & 4.21 & 1.01 \\
China (cn) & 4.63 & 1.54 & Philippines (ph) & 4.11 & 0.97 \\
Pakistan (pk) & 4.46 & 1.02 & Nigeria (ng) & 3.90 & 0.82 \\
Brazil (br) & 4.45 & 1.05 & Indonesia (id) & 3.88 & 1.01 \\
\hline
\end{tabular}
\label{tab:avg_accessibility}
\vspace{-3mm}
\end{table}

\csnote{I recommend you first discuss a simple table/figure in which you report the total/average number of violations per country and how many of these are "critical".}
\mhnote{Replaced the figure with a box plot, I think it is better readable now}


Table~\ref{tab:avg_accessibility} provides an overview of the average number of accessibility and critical issues per website for each country. The table shows widespread accessibility violations across all the countries analyzed, with the average number of critical issues per website showing considerable variation. 
Bangladesh exhibits the highest averages, with 5.48 total issues and 1.29 critical issues per website, indicating substantial accessibility challenges. In contrast, countries like Nigeria (3.90 total issues, 0.82 critical issues) and Indonesia (3.88 total issues, 1.01 critical issues) report lower averages, reflecting relatively better compliance with WCAG guidelines.

\begin{figure}
\centering
\includegraphics[width=.9\linewidth]{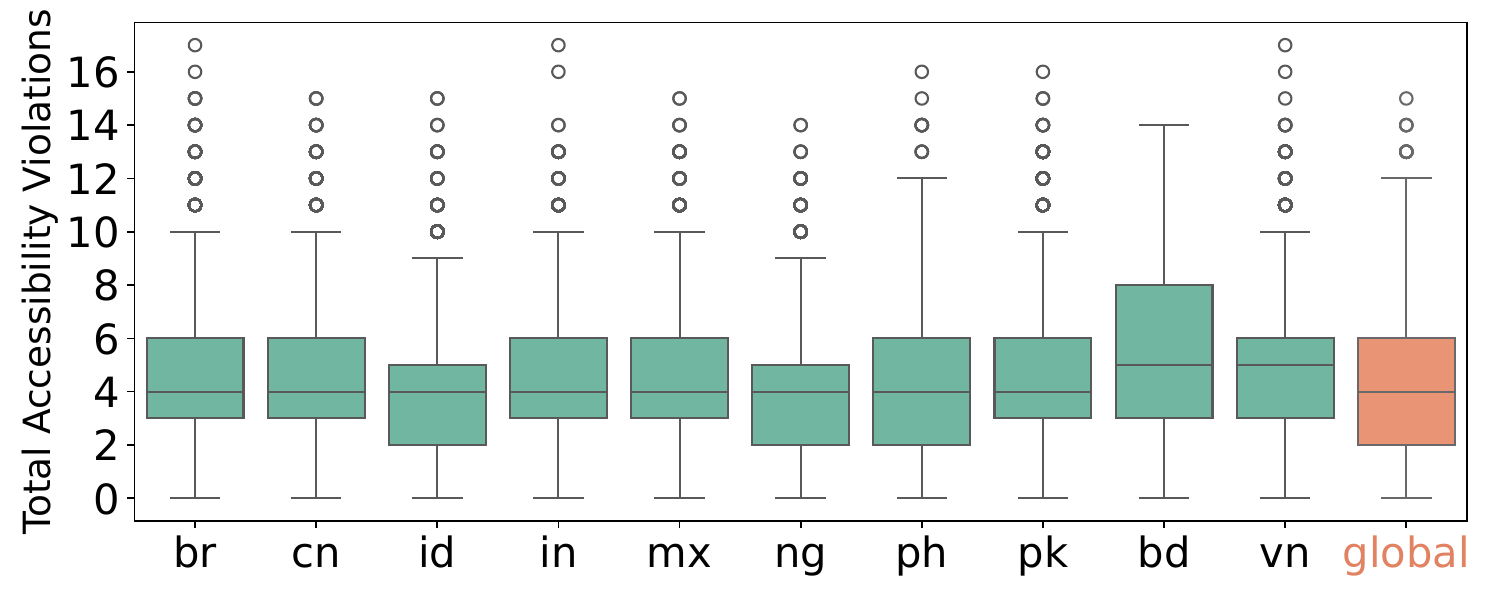}
\vspace{-5mm}
\caption{Distribution of accessibility violations across countries.}
\label{fig:accessibility_violation_distribution}
\vspace{-5mm}
\end{figure}


Figure~\ref{fig:accessibility_violation_distribution} further investigates 
the distribution of accessibility violations across countries, highlighting stark contrasts in \wcag~compliance within the Global South, alongside data for the global top 10k websites as a benchmark. Globally (orange boxplot), 27\% of the top 10k websites exhibit more than five accessibility violations, revealing room for improvement even among leading web platforms. Bangladesh stands out, with nearly 50\% of websites exceeding five accessibility violations. 
This aligns with prior studies~\cite{ahmed2020accessibility, baowaly2012accessibility} that identified significant issues in Bangladeshi websites, particularly in government and education sectors. In contrast, countries like India (30.2\%), the Philippines (25.7\%), and Nigeria (21.9\%) report fewer websites exceeding this threshold, indicating better compliance. Disparities grow at higher thresholds: over 32\% of Bangladeshi websites report more than seven violations, compared to 8\% or less in the Philippines, Nigeria, Mexico, and globally among the top 10k. These discrepancies underscore substantial accessibility gaps in Bangladesh, likely stemming from the lack of robust regulatory frameworks, insufficient incentives to enforce inclusive design practices, and insufficient knowledge or awareness of accessibility laws and guidelines among developers~\cite{baowaly2012accessibility, vinadeunveiling}.

Vietnam also exhibits a significant number of websites affected by accessibility issues, with 38.5\% of its websites exceeding five violations, reflecting challenges in compliance. 
In contrast, countries with established accessibility laws, such as India and Brazil, show lower levels of violations. For example, India’s \textit{Rights of Persons with Disabilities Act, 2016} mandates digital accessibility, fostering greater compliance across public and private sectors. Likewise, Brazil’s \textit{eMAG - e-Government Accessibility Model}, enacted in 2014, supports accessibility guidelines that encourage inclusive design practices and help address gaps in developer knowledge.
\mvnote{would be interesting to go back in time -- wayback machine? -- and see the impact of these acts. Not for now of course :P}


Countries with higher accessibility violations, such as Bangladesh, Pakistan, and Vietnam, often grapple with deeply rooted social stigma surrounding disabilities, which exacerbates the problem. In these societies, individuals with disabilities are frequently marginalized, and their needs are overlooked in public discourse. For instance, in Pakistan, the national census officially reports that only 0.48\% (3.3 million)~\cite{pak_ministry} of its population is disabled, a figure starkly contradicted by the British Council's estimate of 27 million~\cite{pak_british_council}. This significant disparity reflects how stigma surrounding disabilities leads to systemic underreporting and disregard for disability-related issues. This societal stigma not only impacts public perception but also influences government policies and priorities, leading to insufficient investment in accessibility initiatives and digital inclusivity. Similarly, in Bangladesh and Vietnam, societal attitudes often undermine efforts to promote awareness and education on accessibility, further hindering progress~\cite{bd_disable, Vietnam_stigma}.

These findings underscore the critical role of regulatory frameworks, education, and societal attitudes in promoting web accessibility. The stark contrast between countries with high proportions of violations, like Bangladesh and Vietnam, and those with lower proportions, such as Brazil and India, underscores the importance of government policies, national advocacy, and developer training in fostering inclusive digital environments. Addressing societal stigma is equally crucial, as it directly affects the prioritization of accessibility in governance and the creation of a more inclusive digital landscape for individuals with disabilities.

\subsection{Common Accessibility Violations and Their Impact}

\begin{table}
\centering
\setlength{\tabcolsep}{3pt}
\caption{Top 10 Most Common Accessibility Issues}
\vspace{-2mm}
\begin{tabular}{cc|cc}
\hline
\textbf{Issue}            & \textbf{Occurrences} & \textbf{Issue}             & \textbf{Occurrences} \\
\hline
Color Contrast            & 76,040              & Button Name                & 20,640              \\
Link Name                 & 62,440              & Frame Title                & 13,537              \\
Image Alt                 & 43,915              & Label                      & 10,121              \\
Meta-Viewport             & 26,230              & List Structure             & 9,744               \\
HTML Has Language         & 21,571              & Select Name                & 7,955               \\
\hline
\end{tabular}
\label{tab:common_issues}
\vspace{-3mm}
\end{table}

\begin{figure}
\centering
\includegraphics[width=.98\linewidth]{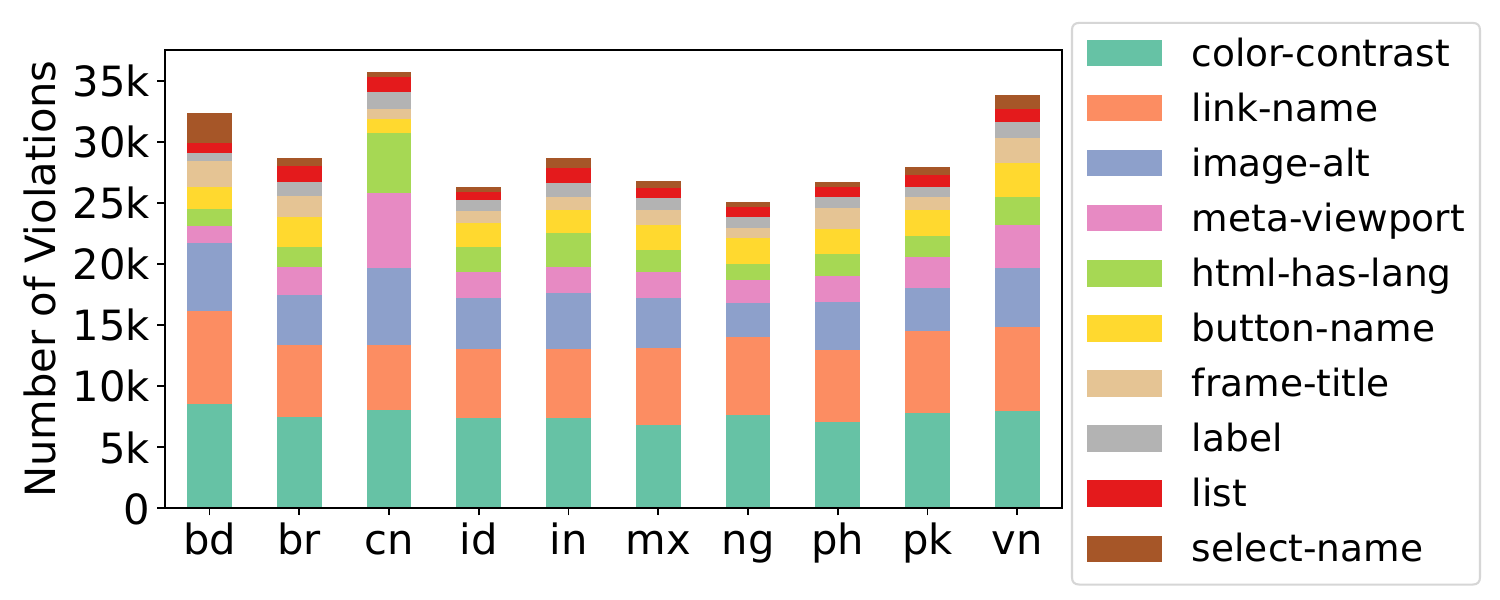}
\vspace{-5mm}
\caption{Number of different accessibility issues identified across countries.}
\label{fig:accessibility_violation_breakdown}
\vspace{-5mm}
\end{figure}

Next, we investigate which specific accessibility challenges, \eg low contrast or lack of alt-text, are responsible for the identified violations.  
Table~\ref{tab:common_issues} shows the top 10  accessibility issues identified across countries, highlighting those that pose the most substantial barriers to web accessibility\mvnote{I don't follow this sentence. Unclear what u are highlighting. Did you do any filtering?}. Figure~\ref{fig:accessibility_violation_breakdown} shows the distributions of the top 10 accessibility issues across countries. Among these, \textit{color contrast} violations are the most frequent. Individuals with low vision often struggle to distinguish elements with poor contrast, as areas of brightness and darkness may blend together, making it difficult to discern outlines, borders, and text details. Text with insufficient luminance contrast against its background is particularly challenging to read, especially for the many people who have low vision or color blindness. Without adequate color contrast, users may find it difficult to read and interact with content. Bangladesh exhibited the highest number of color contrast violations (85.6\%), followed by China (80.3\%) and Vietnam (80.03\%), indicating a significant need to address text visibility and clarity on mobile screens in these regions. 

Link name issue is another widespread problem. Missing or non-descriptive link names create substantial challenges for users relying on screen readers or keyboard navigation. Users with visual impairments and/or 
unable to use a mouse depend on meaningful link descriptions to understand the purpose and destination of each link. Without accessible link names, navigation becomes confusing and inaccessible, as users cannot determine where a link will lead. Again, Bangladesh shows the highest count of link name violations (76\%), followed by Vietnam (68\%) and Pakistan (67\%). 
Image alt text is also missing in a large number of cases; China recorded the most missing alt text cases (63\%), followed by Bangladesh (55\%) and India (46\%). 
Alt text is crucial for conveying an image’s content and purpose, allowing screen readers to convert text into sound or braille. Without it, images become inaccessible, creating gaps in understanding and engagement. 

Meta-viewport configuration issue is also prevalent, affecting how websites scale and display on mobile devices. Proper viewport settings are essential for mobile accessibility, allowing users to zoom and adjust text size as needed. Parameters like \texttt{user-scalable="no"} or restrictive \texttt{maximum-scale} values limit the ability of users with low vision to enlarge content, creating usability challenges on smaller screens. Users with partial vision often rely on zooming functions or screen magnifiers to make text readable, and improper viewport settings can prevent these adjustments. China recorded the most viewport configuration issues (61\%), followed by Vietnam (34\%) and Pakistan (25\%), highlighting the need for mobile optimization to enhance accessibility.

\subsection{Mobile Accessibility Challenges}
\mhnote{The previous Section was for the general web accessibility guidelines violations. My planning for this section was to focus on guidelines which are specific to the mobile web as suggested here https://www.w3.org/TR/mobile-accessibility-mapping/. So, these guidelines are not "new" but rather a subset of the previous sections. As we used the mobile versions of the web pages, the results are also quite similar except for Bangladesh. To avoid reputation I shorted the section}
\mvnote{I am not sure this section has much value. The point is that by restricting to a fewer set of challenges, there is less. Kinda expected? Also it makes me question what is the sense to check for non mobile challenges when we study mobile webpages and mobile devices -- so kinda invalidating the previous analysis. You might want to consider to really really shrink this part and just mention how many of the challenges u have reported before are unique or more important for mobile. Then just focus on the rest which is more interesting, aka impact on blind, visual impaired, etc.}
\begin{figure}
\centering
\includegraphics[width=.9\linewidth]{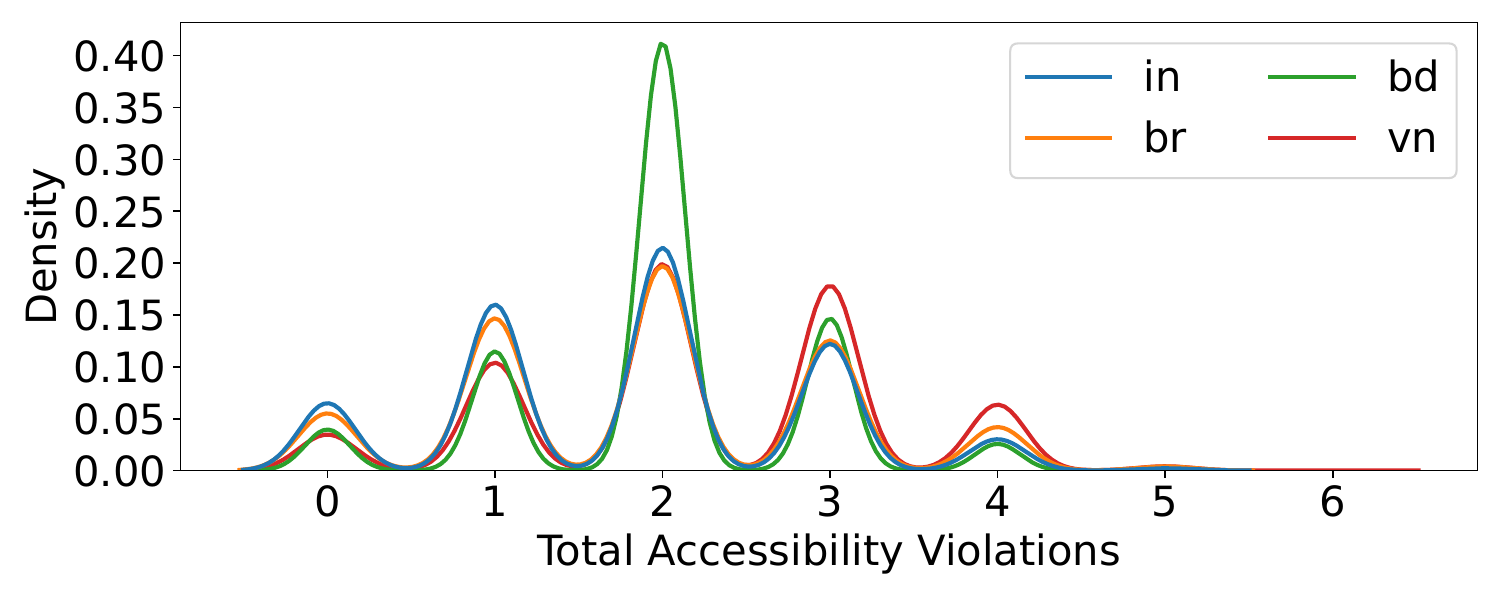}
\vspace{-5mm}
\caption{Distribution of mobile accessibility violations for Bangladesh, Brazil, India, and  Vietnam.}
\label{fig:mobile_accessibility_violation_distribution}
\vspace{-5mm}
\end{figure}

\mvnote{I would just start by saying 2.1 and 2.2 focus on mobile. No need to talk about non-web, since we don't focus on that content}Although \wcagtwo~is highly relevant to both web and non-web mobile content and applications~\cite{wcag_mobile_2015}, the additional guidelines introduced in \wcagtwoone~and \wcagtwotwo~provide more targeted support for mobile accessibility~\cite{wcag_two_two}. These updates address critical elements such as color contrast\mvnote{how is this different from the one u mentioned before?}, touch target sizing, and screen navigability, which are essential for improving the mobile web experience. For example, the \emph{touch-size} guideline ensures that touch targets are sufficiently large and well-spaced, addressing a key challenge of small screens. This is particularly critical for users with fine motor impairments, such as arthritis or tremors, who may struggle with small or tightly packed touch elements. 

We identified 10 
accessibility tests from Lighthouse that closely align with \wcagtwotwo~standards for mobile usability, as shown in Table~\ref{tab:accessibility-checks-mobile}.\mvnote{this table looks weird} 
These tests offer a comprehensive framework for assessing and improving the accessibility of mobile websites, ensuring they meet the needs of diverse users and provide an inclusive experience.

\begin{table}
\centering
\caption{Mobile Accessibility Checks Categorized by WCAG Principles}
\label{tab:accessibility-checks-mobile}
\vspace{-2mm}
\begin{tabular}{|p{0.3\columnwidth}|p{0.6\columnwidth}|}
\hline
\textbf{Principle} & \textbf{Checks} \\ \hline
\textbf{Perceivable} & 
\texttt{color-contrast}, \texttt{meta-viewport}, \texttt{touch-size}, \texttt{input-image-alt} \\ \hline
\textbf{Operable} & 
\texttt{scrollable-region-focusable} \\ \hline
\textbf{Understandable} & 
\texttt{label}, \texttt{link-name} \\ \hline
\textbf{Robust} & 
\texttt{button-name} \\ \hline
\end{tabular}
\vspace{-3mm}
\end{table}

Figure~\ref{fig:mobile_accessibility_violation_distribution} shows the distribution of \textit{mobile accessibility}\mvnote{or maybe mobile-specific? we need a term to separate from before.} violations for four representative countries: Bangladesh, Brazil, India, and  Vietnam. The results for other countries fall between these distributions, highlighting varying levels of compliance. While Bangladesh exhibited a high overall count of accessibility violations (Figure~\ref{fig:accessibility_violation_distribution}), 76.5\% of its websites have fewer than three mobile accessibility violations. This indicates that despite the higher overall violations, most websites adhere relatively well to mobile accessibility guidelines. Similarly, 
73.9\% of the Indian websites 
are affected by three violations. These findings align with~\ref{sec:accessibility_scores_regions}, which highlights higher accessibility compliance in these regions. \mvnote{Fix this sentece. Are these papers also looking into mobile accessibility?} In contrast, only 57\% of the Vietnamese 
websites 
are affected by less than three violations, indicating 
a significant barrier for mobile users. Brazil’s performance fell between these extremes, reflecting moderate levels of compliance and mobile accessibility challenges.

\mvnote{I don't think this paragraph says much} While this analysis suggests that mobile accessibility issues are less concerning than general issues \mvnote{I am not sure this is correct, it is really just that it is a subset}, addressing severe issues such as \texttt{color-contrast}, \texttt{link-name}, \texttt{meta-viewport}, and \texttt{button-name} violations remains critical to improving accessibility and usability. By prioritizing these areas, developers can significantly enhance the mobile web experience, particularly in the Global South, where mobile-first access is essential.


\subsection{Accessibility Challenges for Individuals with Disabilities}

\begin{table}
\centering
\caption{Percentage of websites with no critical accessibility violations for Blind users by country.}
\vspace{-2mm}
\begin{tabular}{cc|cc}
\hline
\textbf{Country} & \textbf{Percentage (\%)} & \textbf{Country} & \textbf{Percentage (\%)} \\
\hline

\texttt{ng} & 53.56 & \texttt{br} & 42.19 \\ \hline
\texttt{pk} & 46.29 & \texttt{in} & 40.74 \\ \hline
\texttt{ph} & 44.56 & \texttt{vn} & 33.53 \\ \hline
\texttt{id} & 43.58 & \texttt{bd} & 31.52 \\ \hline
\texttt{mx} & 43.02 & \texttt{cn} & 29.90 \\ \hline

\end{tabular}
\label{tab:critical_violations_blind}
\vspace{-5mm}
\end{table}

Accessibility challenges impact individuals with various disabilities, including \texttt{blindness}, \texttt{deafblindness}, \texttt{low vision}, \texttt{mobility impairments}, and \texttt{deafness}. Each type of disability requires specific accommodations to ensure equitable access to digital content. For instance, blind users depend on screen readers and well-structured, accessible content, while individuals with low vision rely on features like magnification, high contrast, and text resizing for better readability.
Deafblind users may require advanced assistive technologies like braille displays. Individuals with mobility impairments may face challenges in using a mouse or touchscreen and often depend on alternative input methods such as keyboard navigation, voice commands, switch devices, or adaptive hardware to interact with digital content effectively. Similarly, deaf users depend on subtitles, transcripts, or visual alerts for content comprehension. Critical violations, as defined by Lighthouse and axe-core, are issues that directly prevent users with disabilities from accessing key features or content. These violations demand urgent remediation to ensure inclusivity. While axe-core provides an impact level based on expected severity, the actual impact may vary based on context, underscoring the need for tailored assessments.


\myparagraph{Critical Accessibility Violations for Blind Users.} 
Table~\ref{tab:critical_violations_blind} summarizes 
the critical accessibility violations observed for blind users across the Global South. 
Globally, only 40.89\% of websites meet critical accessibility standards, underscoring substantial barriers for blind users in accessing essential features and content. At the country level, notable variations are evident. Nigeria performs relatively well, with 53.56\% of websites meeting the benchmark, followed by Pakistan (46.29\%) and the Philippines (44.56\%). In contrast, Bangladesh (31.52\%) and China (29.90\%) show the lowest proportions of websites passing the test, highlighting significant accessibility shortcomings. 
For example, \textbf{\texttt{https://fmcabuja.gov.ng}}, the official website of the Federal Medical Centre in Nigeria, is missing alt text for images, lacks proper use of ARIA attributes (e.g., \texttt{aria-allowed-attr} and \texttt{aria-required-children}), and does not have clear button names or labels, making it difficult for a blind person to access the website using a screen reader.
\csnote{Give examples of websites with a high number of such violations and explain why it is bad. Consider adding such examples for the next two paragraphs as well}\mhnote{done}

\begin{figure}
\centering
\includegraphics[width=.9\linewidth]{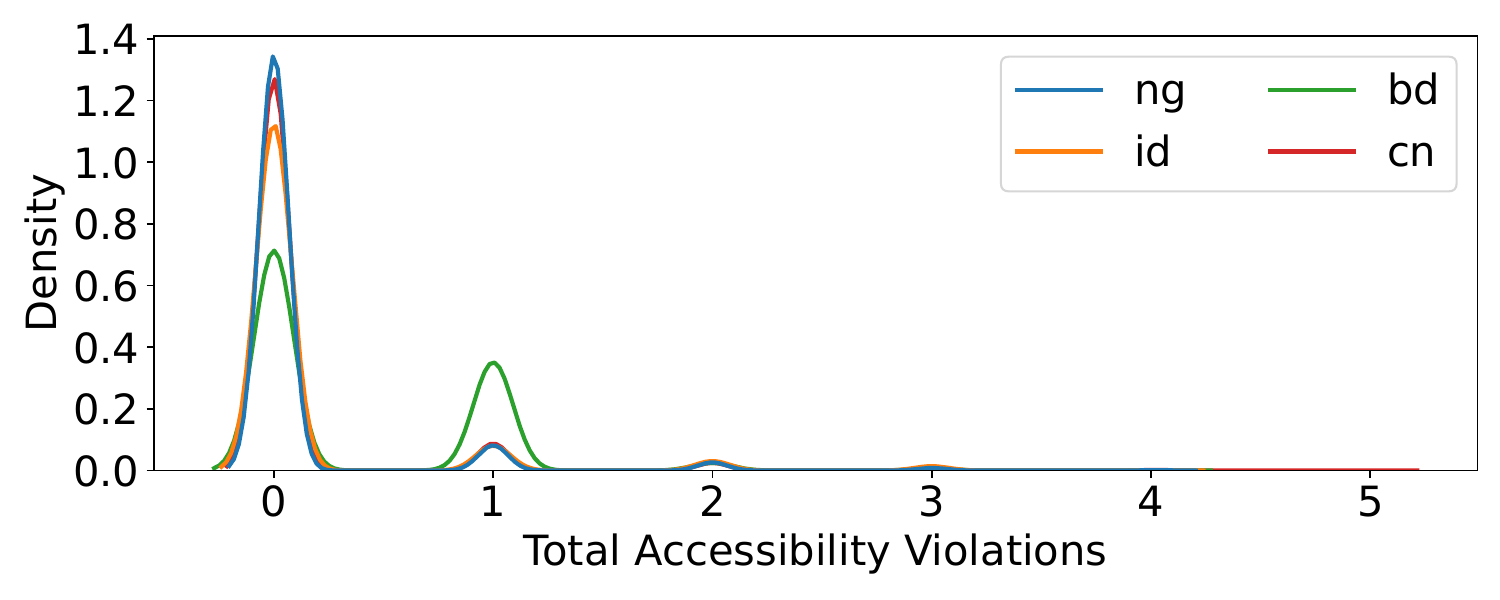}
\vspace{-5mm}
\caption{Distribution of critical accessibility violations for users with mobility impairments across countries.}
\label{fig:critical_violations_mobility}
\vspace{-6mm}
\end{figure}

\myparagraph{Critical Accessibility Violations for Mobility Impairments Users.} 
Figure~\ref{fig:critical_violations_mobility} shows four representative distributions (China, Nigeria, Indonesia, and Bangladesh) of critical accessibility violations affecting users with mobility impairments. 
Compared to blindness, the figure shows much higher accessibility support, 
with 86.7\% of the overall websites meeting the critical accessibility criteria. Nigeria leads in accessibility for mobility disabilities (91.91\%), closely followed by China (91.15\%) and Indonesia (89.88\%).
Bangladesh lags significantly, with only 65.01\% of websites meeting the critical criteria, indicating a substantial gap. For instance, \textbf{\texttt{http://krishi.gov.bd}}, a website of the Ministry of Agriculture in Bangladesh, lacks proper keyboard navigation and does not implement accessible features such as well-structured ARIA attributes. Additional issues, including improper button size (\texttt{target-size}), inadequate logical tab order, and missing skip links, make navigation difficult for individuals with mobility impairments. 

\begin{table}
\centering
\caption{Percentage of websites with no critical accessibility violations for low vision users by country.}
\vspace{-2mm}
\begin{tabular}{cc|cc}
\hline
\textbf{Country} & \textbf{Percentage (\%)} & \textbf{Country} & \textbf{Percentage (\%)} \\
\hline
\texttt{bd} & 85.74 & \texttt{id} & 78.26 \\ \hline
\texttt{ng} & 81.50 & \texttt{br} & 77.01 \\ \hline
\texttt{mx} & 79.14 & \texttt{pk} & 74.95 \\ \hline
\texttt{in} & 78.84 & \texttt{vn} & 65.01 \\ \hline
\texttt{ph} & 78.72 & \texttt{cn} & 38.52 \\ \hline
\end{tabular}
\label{tab:country_percentages}
\vspace{-5mm}
\end{table}

\myparagraph{Critical Accessibility Violations for Low Vision Users.} Table~\ref{tab:country_percentages} shows the accessibility assessment for users with low vision across the Global South; overall, 73.77\% of the websites meet the criteria, thus outperforming blindness support while under-performing support for mobility impairments. 
Notably, Bangladesh performs better in this category compared to its results for other disabilities, with 85.74\% of websites meeting the benchmark. On the other hand, Chinese websites remain a concern, with only 38.52\% meeting the criteria, highlighting persistent challenges for low vision accessibility. For instance, \textbf{\texttt{https://www.mfa.gov.cn}}, the official website of the Ministry of Foreign Affairs in China, lacks meta-viewport settings and sufficient color contrast, making it difficult for low vision users to navigate. Other countries, such as Nigeria (81.50\%), Mexico (79.14\%), and the Philippines (78.72\%), perform close to the overall average, while Vietnam (65.01\%) lags slightly behind.
\csnote{Discuss example of Chinese or Russian websites with poor accessibility for this category of users.}\mhnote{done}

\section{Conclusion}
This study provides the first large-scale analysis of mobile web accessibility across the Global South, evaluating 100,000 websites from 10 countries. Our findings reveal the uneven impact of accessibility issues on different user groups. While websites show better compliance for users with mobility impairments or low vision, blind users face critical violations like missing link names and alt text. This disparity highlights the urgent need for targeted interventions to address the unique challenges faced by blind users. 
Another notable finding is the correlation between national laws, the social stigma surrounding disabilities, and accessibility compliance. Countries with robust legislation, such as India’s \textit{Rights of Persons with Disabilities Act} and Brazil’s \textit{eMAG - e-Government Accessibility Model}, demonstrate higher adherence to accessibility standards. In contrast, countries like Bangladesh, Pakistan, and Vietnam, where stigma about disabilities is prevalent, show significantly lower compliance. This highlights how societal attitudes can hinder both policy development and enforcement. Further, this underscores the pivotal role of governance and policy in fostering inclusive digital environments.  Despite these insights, there remains considerable room for improvement across all regions. Addressing widespread violations, especially those affecting the most vulnerable groups, will require stronger regulatory frameworks, developer education, and widespread adoption of accessibility-first design practices. These efforts are essential to bridge the accessibility gap and promote digital equity in the Global South.  
\bibliographystyle{ieeetr}
\bibliography{reference}

\end{document}